\documentclass[useAMS,usenatbib,usegraphicx]{mn2e}

\newcommand{\taut}{\tau_{\rm T}}

\title[Reflection Models]
  {A comprehensive range of X-ray ionized reflection models} 
\author[R.R. Ross \& A.C. Fabian]
  {R.~R.~Ross$^1$\thanks{rross@holycross.edu} and A.~C.~Fabian$^2$\\
$^1$Physics Department, College of the Holy Cross, Worcester, MA 01610, USA \\
   $^2$Institute of Astronomy, Madingley Road, Cambridge CB3 0HA}

\pagerange{\pageref{firstpage}--\pageref{lastpage}}
\pubyear{2003}

\usepackage{times}

\begin{document}

\label{firstpage}

\maketitle

\begin{abstract}
X-ray ionized reflection occurs when a surface is irradiated with
X-rays so intense that its ionization state is determined by the
ionization parameter $\xi\propto F/n$, where $F$ is the incident flux
and $n$ the gas density. It occurs in accretion, onto compact objects
including black holes in both active galaxies and stellar-mass
binaries, and possibly in gamma-ray bursts. Computation of model
reflection spectra is often time-consuming. Here we present the
results from a comprehensive grid of models computed with our code,
which has now been extended to include what we consider to be all
energetically-important ionization states and transitions. This grid
is being made available as an ionized-reflection model, {\sc reflion},
for XSPEC.

\end{abstract}

\begin{keywords}
X-rays: general -- accretion, accretion discs -- galaxies: active -- 
radiative transfer -- line: formation
\end{keywords}

\section{Introduction}
The process of backscattering and fluorescence of X-rays from cosmic
matter known as X-ray reflection is commonly observed throughout
astronomy. We are concerned here with the situation where the X-ray
irradiation is so intense that it determines the ionization state of
the material. This occurs in accreting X-ray sources such as Galactic
Black Hole Candidates (BHC) and Active Galactic Nuclei (AGN), and may
also be relevant to Gamma-Ray Bursts (GRB).

The irradiated gas is Compton thick, which requires that
Comptonization of the radiation as well as the necessary atomic
physics be dealt with. Results of such computations have been
published over the last ten years, first for constant-density 
atmospheres (Ross \& Fabian 1993; \.{Z}ycki et al.\ 1994; 
Ross, Fabian \& Young 1999) and more recently for density structures
in hydrostatic equilibium (Nayakshin, Kazanas \& Kallman 2000;
Ballantyne, Ross \& Fabian 2001; R\'{o}\.{z}a\'{n}ska et al.\ 2002; 
Mauche et al.\ 2004). Reasonable agreement is found between most of these 
results (P\'{e}quignot et al.\ 2002).

Most methods are computationally very time consuming, so only limited
regions of parameter space have been explored. Our method runs
relatively rapidly, on the other hand, when the illuminated gas is 
assumed to have constant density. Such models are still useful for several
reasons. The overall geometry is still not understood for many sources.
Even when a simple accretion-disc structure is assumed, detailed
calculations of hydrostatic equilibrium are subject to uncertainties
in the total thickness of the disc (and hence the vertical component of
the local gravitational field), the boundary condition (either pressure
or height) at the base of the illuminated layer, the external pressure
exerted by the corona, the geometry of the illumination, the influence of 
magnetic fields, etc. Furthermore, Ballantyne et al.\ (2001) found that 
models for reflection by atmospheres in hydrostatic equilibrium could be 
fit by diluted versions of constant-density reflection models.

We have now assembled an extended grid of
models for which the ionization parameter, irradiating spectral index
and iron abundance are free parameters. At the same time the original
code has been expanded to include further ionization species. All
abundant species and their important transitions are now included. The
purpose of this paper is to present example spectra from across this
grid which can be used as a guide to the interpretation of AGN, BHC
and GRB X-ray spectra.

\section{Method}
\subsection{Radiative Transfer}

Our calculations extend and improve on the models discussed by Ross, Fabian
\& Young (1999). Now the illuminating radiation is assumed to have a cutoff 
power-law spectrum,
\begin{equation}
F_E = AE^{-\Gamma + 1}\exp(-E/E_{\rm c}),
\end{equation}
that extends to higher photon energies. Different values of the photon index 
$\Gamma$ are treated, while the cutoff energy is fixed at 
$E_{\rm c}=300{\rm\ keV}$. The amplitude $A$ is chosen so that the total 
illuminating flux, $F_{\rm tot}=\int F_E\,dE$, corresponds to a desired value 
of the ionization parameter,
\begin{equation}
\xi = \frac{4\pi F_{\rm tot}}{n_{\rm H}}.
\end{equation}

The illumination is incident on the upper surface of a plane-parallel slab
representing the top layer of an optically thick medium such as an accretion 
disc. The slab has hydrogen number density $n_{\rm H} = 10^{15}{\rm\ cm}^{-3}$ 
and Thomson depth $\taut\geq 5$. (Larger Thomson depths are treated for 
larger values of $\xi$, since the illumination can penetrate to 
greater depths when the gas is more highly ionized.) No net flux is allowed 
to enter the slab from below, so that the emergent radiation is due entirely
to reprocessed illumination (``reflection''). Additional radiation from 
beneath the surface layer could be approximated by adding a soft blackbody 
spectrum to the calculated reflection spectrum (see Section 4).

As described by Ross \& Fabian (1993), the penetration of the illuminating 
radiation is treated separately from the transfer of the diffuse radiation 
produced via scattering or emission within the gas. In order to extend the 
treatment of the diffuse radiation to photon energies above 100 keV, Compton 
scattering is treated using the Fokker-Planck equation of Cooper (1971):
\begin{equation}
\left(\frac{\partial n}{\partial t}\right)_{\rm FP}=
\frac{n_{\rm e}\sigma_{\rm T}}{m_{\rm e}c}\frac{1}{E^2}
\frac{\partial}{\partial E}
\left[\alpha(E,\theta)E^4
\left(n+\theta\frac{\partial n}{\partial E}\right)\right],
\end{equation}
where $n$ is the photon occupation number, $E$ is the photon energy,
$n_{\rm e}$ is the free-electron number density, and $\theta=kT$.  
Here the Fokker-Planck equation of Kompaneets (1957) has been 
modified by the term
\begin{equation}
\alpha(E,\theta)=\frac{1+\beta(\theta)/(1+0.02E)}
{1+9\times 10^{-3}E+4.2\times 10^{-6}E^2}\,,
\end{equation}
with $E$ in keV and with
\begin{equation}
\beta(\theta)=\frac{5}{2}\frac{\theta}{m_{\rm e}c^2}
+\frac{15}{8}\left(\frac{\theta}{m_{\rm e}c^2}\right)^2
\left(1-\frac{\theta}{m_{\rm e}c^2}\right).
\end{equation}
Equation~(3) is applicable for $E\la 1{\rm\ MeV}$.  
The diffuse radiation field is taken to satisfy the steady-state 
Fokker-Planck/diffusion equation,
\begin{equation}
\left(\frac{\partial n}{\partial t}\right)_{\rm FP}
+\frac{\partial}{\partial z}\left(\frac{c}{3\kappa}
\frac{\partial n}{\partial z}\right)
+\frac{j_Eh^3c^3}{8\pi E^3} - c\kappa_{\rm A}n \equiv 0\,,
\end{equation}
where $z$ is the vertical height within the slab, $j_E$ is the spectral 
emissivity, $\kappa=\kappa_{\rm A}+\kappa_{\rm KN}$ is the total opacity 
(per volume), $\kappa_{\rm A}$ is the absorption opacity, and 
$\kappa_{\rm KN}$ is the Klein-Nishina opacity for Compton scattering.  
We apply equation~(6) to radiation with $10^{-3}<E<10^3{\rm\ keV}$.  
The energy resolution is $E/{\Delta E}=50$ throughout the 0.1--10 keV 
spectral range (except for a resolution of 70 around the iron K$\alpha$
lines), with somewhat lower resolution outside that range.

\subsection{Atomic Data}

As the radiation field is relaxed to a steady state, the local temperature 
and fractional ionization of the gas are found by solving the equations of 
thermal and ionization equilibrium, as described by Ross \& Fabian (1993).
For this non-LTE calculation,  
we have extended the number of elements and the range of ions 
treated, and we have updated much of the atomic data employed.  In addition 
to fully-ionized species, the following ions are included in the calculations:
C\,{\sc iii--vi}, N\,{\sc iii--vii}, O\,{\sc iii--viii}, Ne\,{\sc iii--x}, 
Mg\,{\sc iii--xii}, Si\,{\sc iv--xiv}, S\,{\sc iv--xvi}, and 
Fe\,{\sc vi--xxvi}.

Photoionization cross sections for all subshells of the ions
treated are calculated from the fits of Verner \& Yakovlev (1995).  
Rates for direct collisional ionization and excitation
autoionization are taken from Arnaud \& Rothenflug (1985) and
Arnaud \& Raymond (1992).  For carbon through sulfur, Auger
ionization is treated as described by Weisheit \& Dalgarno (1972).
For iron, we employ the probabilities for multiple Auger ionization 
calculated by Kaastra \& Mewe (1993), and a generalization of the 
method of Weisheit (1974) is used to calculate the resulting effect 
on the ionization balance.

For all elements except iron, total recombination rates as functions
of density and temperature are derived from the tables of Summers 
(1974).  These rates include two effects that are important at high 
densities: the increase in the recombination rate at low temperatures 
due to three-body recombination and the reduction in the dielectronic 
recombination rate at high temperatures due to collisional ionization 
of the highly excited states that follow radiationless recombination 
(Burgess \& Summers 1969).  At low densities, these rates are in
reasonable agreement with more modern fits for combined radiative and 
dielectronic recombination (Arnaud \& Rothenflug 1985).
For iron, rates for radiative and dielectronic recombination are 
calculated using the fits of Arnaud \& Rothenflug (1985) and Verner 
\& Ferland (1996).  Three-body recombination of iron is neglected, 
since it is unimportant at the densities under consideration (Jacobs 
et al.\ 1977).  Rates for radiative recombination directly to 
ground levels are calculated using the Milne relation (e.g., Bates
\& Dalgarno 1962).

The most important emission lines for each ion are treated
in the calculations. 
{ Since the illuminated atmosphere is thick and dense, resonance 
lines are assumed to be optically thick (``case B''), greatly reducing
the number of important lines. A total of 274 multiplets are included;
individual lines within a multiplet are not resolved.}
Line energies and oscillator strengths for resonance lines are taken
from the results of the Opacity Project (Verner, Verner \& Ferland 1996).
Excitation and line energies of nonresonance lines, as well as collision
strengths for all lines, are derived from the CHIANTI database (Dere et
al.\ 2001).  Parameter values missing from these databases are taken
from Kato (1976), Raymond \& Smith (1977), Gaetz \& Salpeter (1983), or 
Landini \& Monsignori Fossi (1990), as necessary. The iron K$\alpha$ lines
treated are the recombination lines of Fe\,{\sc xxvi} and Fe\,{\sc xxv} 
(near 7.0 and 6.7 keV, respectively) and the fluorescence lines of
Fe\,{\sc vi--xvi} (near 6.4 keV). K$\alpha$ fluorescence of
Fe\,{\sc xvii--xxii} is assumed to be suppressed by autoionization (Auger 
effect) during resonance trapping (Ross, Fabian \& Brandt 1996).
Recombinations to excited levels of Fe\,{\sc xvii--xxii} are assumed
to result ultimately in $3{\rm s}\rightarrow 2{\rm p}$ line emission
(Liedahl et al.\ 1990).

Resonance lines require special attention. Resonance-line photons can 
avoid destruction during resonance trapping by two mechanisms: Compton 
scattering out of the narrow line core between resonance scatterings 
and reemission in the far line wings during resonance scattering. The 
fraction avoiding destruction is taken to be
\begin{equation}
f = \frac{(\kappa_{\rm T}/\kappa)P_{\rm cont}+P_{\rm esc}}
    {P_{\rm cont}+P_{\rm coll}+P_{\rm esc}}.
\end{equation}
Here $\kappa_{\rm T}$ is the Thomson opacity, $\kappa$ is the total continuum 
opacity at the line energy, $P_{\rm cont}$ is the probability per resonance 
scattering that the photon undergoes a continuum process (absorption or Compton
scattering), and $P_{\rm coll}$ is the probability per resonance scattering 
that collisional deexcitation occurs instead of reemission (Hummer 1968). The 
escape probability $P_{\rm esc}$ due to reemission in the far line wings 
(Hummer \& Rybicki 1971) is calculated using the approximation for the $K_2$ 
function given by Hollenbach \& McKee (1979). A Doppler line profile is assumed
for all resonance lines except the Fe K$\alpha$ line, where a Lorentz profile is
assumed due to the importance of the damping wings. Line photons that avoid 
destruction during resonance trapping are added to the continuum by means of 
the local emissivity.

Elements lighter than iron are assumed to have solar abundances
(Morrison \& McCammon 1983), while the abundance of iron is set to a
factor $A_{\rm Fe}$ times its solar value. Models are generated with
underabundances as well as overabundances of iron. On the one hand the
X-ray spectra of some Seyfert galaxies appear to require high
abundance of iron, for example MCG--6-30-15 (Fabian et al.\ 2002) and
1H\,0707-49 (Boller et al.\ 2002), and Shemmer et al.\ (2004) have
reported a strong correlation between metallicity and accretion rate
in a sample of luminous quasars. On the other hand, Nomoto et al.\ 
(1997) have found that Type II supernovae produce the following
average abundances relative to solar values: $A_{\rm C}=0.2$, $A_{\rm
N}=0.01$, $A_{\rm O}=1.3$, $A_{\rm Ne}=0.9$, $A_{\rm Mg}=1.2$, $A_{\rm
Si}=1.1$, $A_{\rm S}=0.7$, and $A_{\rm Fe}=0.35$. In that case,
oxygen, neon, magnesium and silicon are near solar abundances, while
the abundance of iron is markedly reduced.

\subsection{Test Case: an Isolated Slab}

\begin{figure}
\includegraphics[scale=0.35,angle=270]{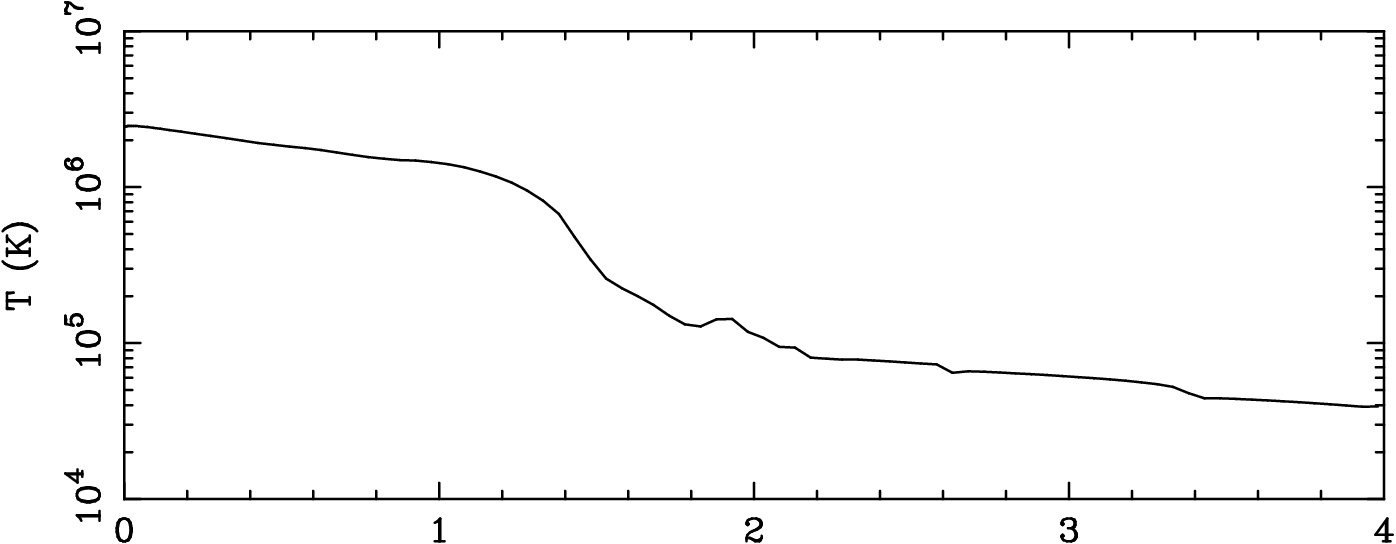}
\vskip0.25cm
\includegraphics[scale=0.35,angle=270]{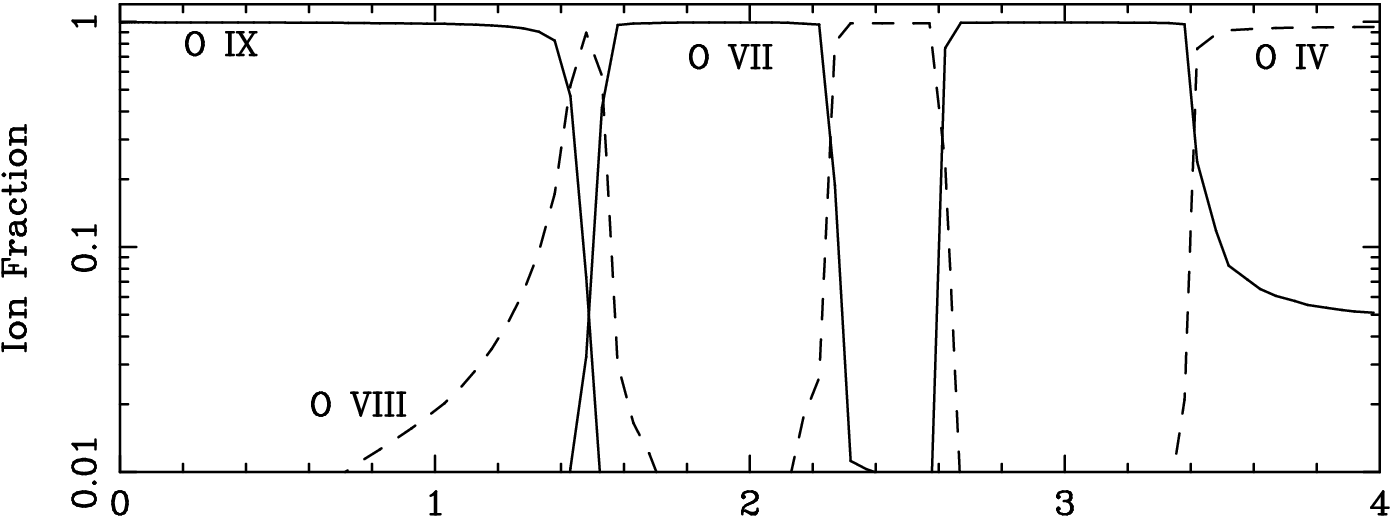}
\vskip0.25cm
\includegraphics[scale=0.35,angle=270]{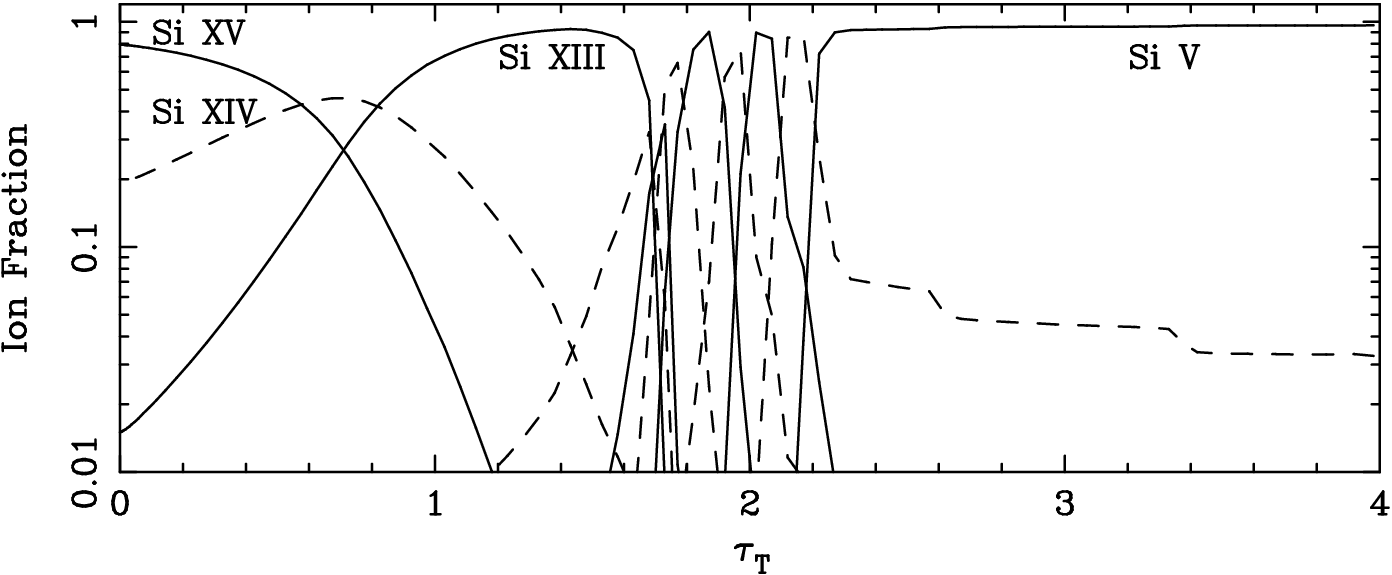}
\caption{Structure as a function of Thomson depth for the {\it isolated} 
slab illuminated on one side. The upper panel shows the temperature,
while the middle and bottom panels show the ion fractions of oxygen
and silicon, respectively.}
\end{figure}

Recently Dumont et al.\ (2003) have criticized calculations, such as ours
and those of Nayakshin et al.\ (2000), that employ the escape 
probability in treating emission lines.
{ Dumont et al.\ use an ``accelerated lambda iteration'' method that includes
a detailed treatment of radiative transfer within the spectral lines themselves.
They point out that their code has the drawback of being time-consuming, which
precludes it from easily generating large grids of models for fitting data. 
In particular, they have applied their method to an {\it isolated} slab that is 
illuminated on one side. The total Thomson depth of the slab is $\taut = 4$, and 
the illuminating flux corresponds to $\xi = 10^3$ erg\,cm\,s$^{-1}$.} 
As a test of our method, we have performed our own computation for this situation. 
We cannot expect complete agreement because our calculations assume that 
helium is always fully ionized. Normally our interest is in reflection by 
a highly optically-thick medium. In order for all of the incident energy 
to escape back out the illuminated surface, the total radiative energy 
density increases with Thomson depth to such a degree that the assumption 
that helium is fully ionized should be satisfactory. When radiation is 
free to leave the far surface of a marginally thick slab, however, helium 
need not be fully ionized throughout. Dumont et al.\ (2003) found a large 
fraction of He$^+$ ions at depths $\taut\geq 3$. Our calculation of
the isolated slab serves mainly as a test for the Lyman-$\alpha$ lines of 
hydrogen-like ions, which are produced in the hotter layers nearer the 
illuminated surface.

\begin{figure}
\includegraphics[scale=0.35,angle=270]{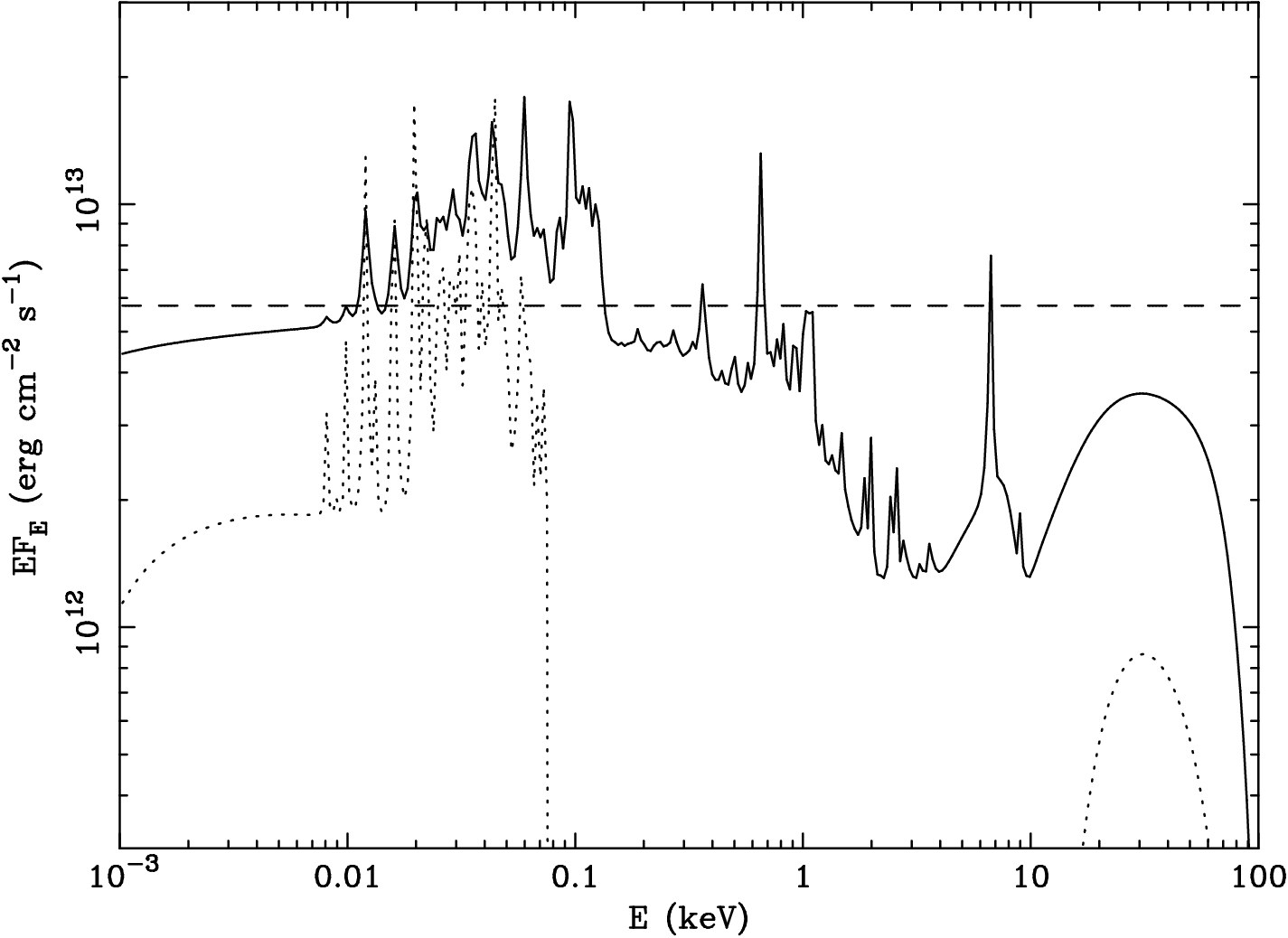}
\vskip0.25cm
\includegraphics[scale=0.35,angle=270]{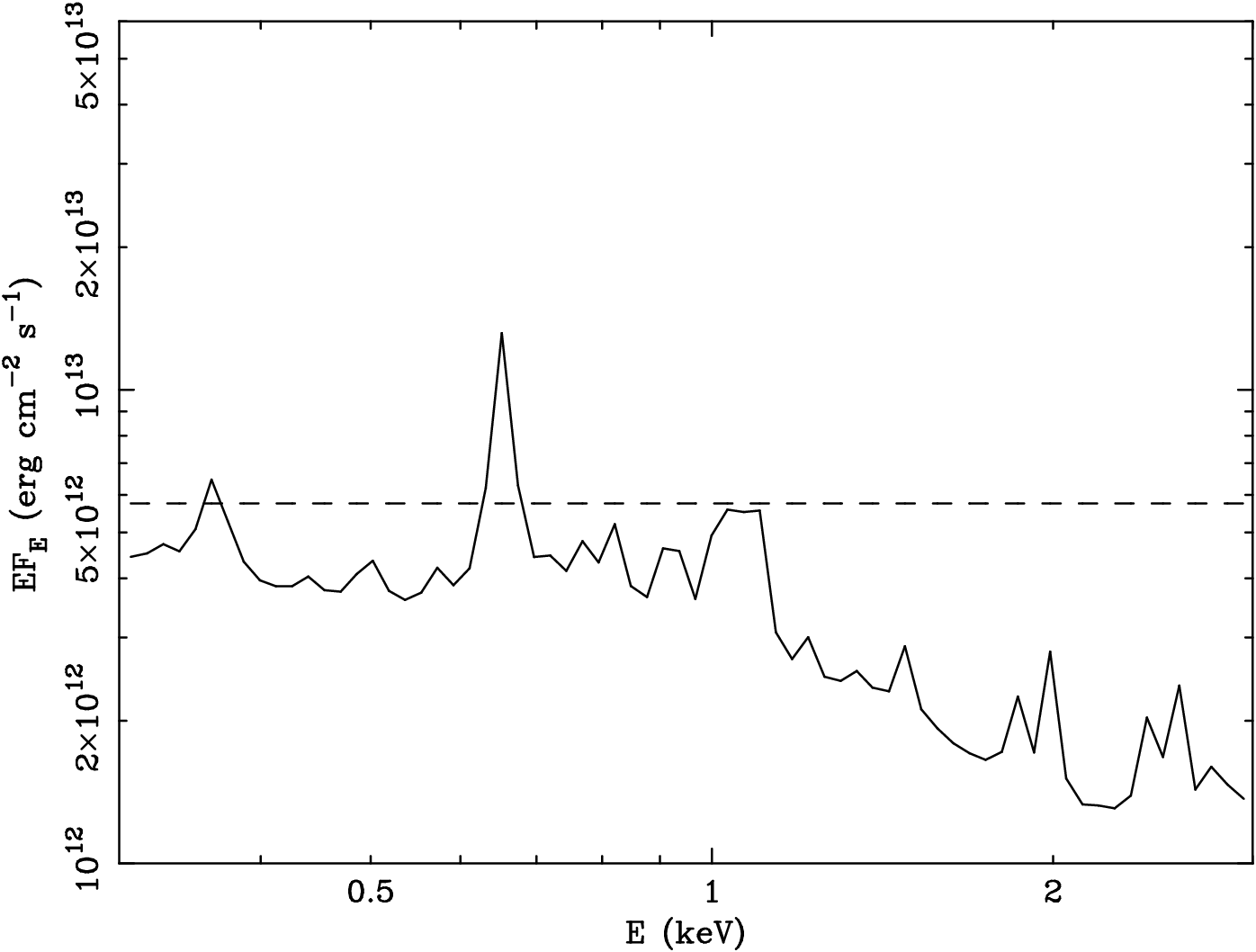}
\caption{Spectra for the {\it isolated} slab illuminated on one side.
The dashed curve is the illuminating spectrum, the solid curve is the 
``reflected'' spectrum emerging from the illuminated side, and the dotted 
curve is the ``transmitted'' spectrum emerging from the far side. The
lower panel shows a close-up of a portion of the broader spectrum shown 
in the upper panel. The spectral resolution is 30.}
\end{figure}

The biggest change required for this model is in the boundary condition
at the far side of the slab. Diffuse radiation is allowed to emerge there,
with the escaping flux calculated in the same way as at the illuminated 
surface (see Foster, Ross \& Fabian 1986). Other changes must also be
made to the method described previously. The gas density is reduced to
$n_{\rm H} = 10^{12}{\rm\ cm}^{-3}$. The illuminating spectrum is a simple
power law with photon index $\Gamma=2$ extending from 0.1~eV to 100~keV,
and the energy resolution is fixed at $E/{\Delta E}=30$. The metal 
abundances are set to the values of Allen (1973), which differ slightly
from those of Morrison \& McCammon (1983).

Figure 1 shows the temperature structure and the ion fractions for oxygen 
and silicon that we calculate for the isolated slab. For $\taut\la 2.5$, 
our results are in good agreement with those of Dumont et al.\ (2003). The
differences at greater Thomson depth are presumably due to the lack of
He$^+$ in our model.

The reflected spectrum that we calculate for the isolated slab is shown in 
Figure~2. Dumont et al.\ (2003) concluded that use of the escape probability
must result in a drastic overestimation of the strength of the Lyman-$\alpha$
lines of hydrogen-like ions. However, we find the Lyman-$\alpha$ lines of
C~{\sc vi} (0.37 keV), O~{\sc viii} (0.65 keV), Mg~{\sc xii} (1.5 keV),
Si~{\sc xiv} (2.0 keV), and S~{\sc xvi} (2.6 keV) to be in excellent
agreement with their results. 
{ In particular, the O~{\sc viii} line has an equivalent width of 27~eV 
with respect to the total (incident plus reflected) continuum, compared
to their value of 28~eV. For the Si~{\sc xiv} line, which is complicated
by a steeper underlying continuum and blending with the He-like line, we 
estimate an EW of 14~eV compared to their value of 9.5~eV.}
In our calculations, Compton scattering of
resonance-line photons out of the narrow line core turns out to be much more
important than reemission in the far line wings. That is, the first term of
the numerator in equation~(7) completely dominates over the second term.

The Fe K$\alpha$ line is somewhat stronger in our calculation. Since it is 
dominated by the Fe~{\sc xxv} intercombination line (6.7 keV), however, this 
is probably due to different approximations in treating the atomic physics. 
Curiously, we find the Lyman-$\alpha$ line of N~{\sc vii} (0.50 keV) to be 
weaker than found by Dumont et al.\ (2003). Differences can be seen in other
emission lines, but these are produced deeper within the slab, where He$^+$
could be a factor. Fig.~2 also shows the calculated spectrum emerging from 
the far side of the slab. Because of the lack of He$^+$, the EUV portion of
the transmitted spectrum cuts off at a higher energy than found by Dumont
et al.\ (2003).

Overall, our method appears to give reasonably accurate results, with 
computing times short enough to allow the production of a large grid of 
reflection models for fitting to observed X-ray spectra.

\section{Results}

For an optically-thick medium, we have calculated model reflection spectra 
covering a range of values for the ionization parameter ($\xi=30$, 100, 300, 
1\,000, 3\,000, and 10\,000 erg\,cm\,s$^{-1}$), the photon index 
($\Gamma=1.0$ to 3.0 in steps of 0.2), and the iron abundance 
($A_{\rm Fe}=0.1$, 0.2, 0.5, 1.0, 2.0, 5.0, and 10.0) relative to its solar value. 
The result is a grid of 462 reflection models.

\begin{figure}
\includegraphics[scale=0.35,angle=270]{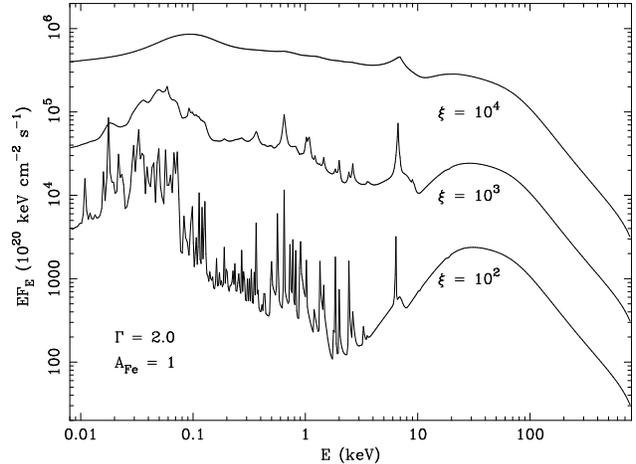}
\caption{Reflected spectra for three values of the ionization parameter,
with $\xi = 10^2$ (bottom curve), $10^3$, and $10^4$ erg\,cm\,s$^{-1}$
(top curve). The incident spectrum has $\Gamma = 2.0$, and iron has solar
abundance.}
\end{figure}

Of course, the ionization parameter has a marked effect on emission and 
absorption features in the reflected spectrum. Figure~3 shows reflected 
spectra for three different values of $\xi$. In each case, the incident 
spectrum has $\Gamma = 2.0$, and iron has solar abundance 
($A_{\rm Fe} = 1$). For $\xi=10^4$ erg\,cm\,s$^{-1}$, the
surface layer is highly ionized, with Fe\,{\sc xxv--xxvii} dominating
for $\taut\la 7$. The only important emission feature is the highly
Compton-broadened iron K$\alpha$ line peaking at 7.0~keV.
For $\xi=10^3$ erg\,cm\,s$^{-1}$, the illuminated gas is
not as highly ionized. Fe\,{\sc xxv} dominates for $\taut\la 1$, and
the lighter elements are no longer fully ionized below that. The strong
iron K$\alpha$ emission is dominated by the Fe\,{\sc xxv} intercombination
line, and Compton broadening is still important. The spectral region
between 0.3 and 3 keV shows K$\alpha$ lines of C\,{\sc vi}, O\,{\sc viii}, 
Ne\,{\sc x}, Mg\,{\sc xii}, Si\,{\sc xiii--xiv}, and S\,{\sc xv--xvi},
as well as a few Fe L$\alpha$ lines, all atop a shallow absorption trough.
The N\,{\sc vii} K$\alpha$ line, at an energy (0.50 keV) just above the
C\,{\sc vi} K-edge, is very weak. 
When $\xi$ is reduced to $10^2$ erg\,cm\,s$^{-1}$, there is very little 
ionization for $\taut\ga\frac{1}{2}$. The narrow Fe K$\alpha$ line near 6.4 keV
is due to fluorescence, and the spectral region below 3 keV exhibits a 
myriad of emission features atop a deep absorption trough.

\begin{figure}
\includegraphics[scale=0.35,angle=270]{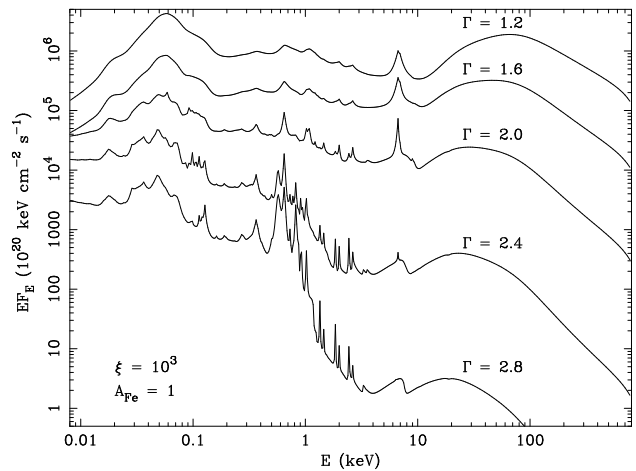}
\caption{Reflected spectra for five values of the photon index,
with $\Gamma = 1.2$ (top curve), $1.6$, $2.0$, $2.4$, and $2.8$ (bottom 
curve). The incident spectrum has $\xi = 10^3$ erg\,cm\,s$^{-1}$, and iron 
has solar abundance. Successive spectra have been offset by factors
of five for clearer presentation.}
\end{figure}

In addition to affecting the overall slope of the reflected spectrum, 
the photon index $\Gamma$ also affects the emission and absorption features, 
since harder illuminating spectra have greater ionizing power. Figure~4 
shows reflected spectra for five different values of $\Gamma$. 
In each case, the incident spectrum has $\xi = 10^3$ erg\,cm\,s$^{-1}$, 
and iron has solar abundance ($A_{\rm Fe} = 1$). 
{ For $\Gamma=1.2$ or $1.6$, the illuminated gas is more highly ionized 
than for $\Gamma=2.0$. The iron K$\alpha$ feature is a broader combination 
of Fe~{\sc xxv} and Fe~{\sc xxvi} lines, and the lines of the lighter 
elements are weaker. With $\Gamma=2.4$ or $2.8$, on the other hand, the 
illuminated gas is less highly ionized. Fe~{\sc xviii}--{\sc xxii} dominate 
at the outer surface instead of Fe~{\sc xxv}, so the Fe K$\alpha$ line is 
much weaker, while the Fe L$\alpha$ lines are stronger and more numerous. 
The K$\alpha$ lines of the lighter elements are stronger (as is the continuum 
absorption), with He-like exceeding H-like for Mg, Si and S. The oxygen lines 
are especially enhanced by a soft illuminating spectrum.}

\begin{figure}
\includegraphics[scale=0.35,angle=270]{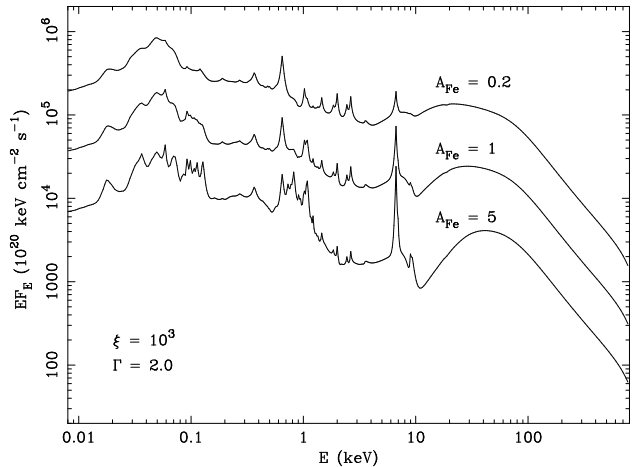}
\caption{Reflected spectra for three values of the iron abundance,
with $A_{\rm Fe}=0.2$ (top curve), 1, and 5 (bottom curve) times solar 
abundance. The incident spectrum has $\xi=10^3$ erg\,cm\,s$^{-1}$
and $\Gamma = 2.0$. Successive spectra have been offset by
factors of five for clearer presentation.}
\end{figure}

The effects of altering the abundance of iron are shown in Figure~5.
Again, the illuminating radiation has $\xi=10^3$ erg\,cm\,s$^{-1}$ and 
$\Gamma=2.0$. With iron reduced to 0.2 times its solar abundance, both
the K$\alpha$ emission line (6.7 keV) and the K-absorption feature
($\sim\!10\,{\rm keV}$) are reduced compared to the reflected spectrum 
with solar abundance. Also, less Fe L$\alpha$ emission is blended with 
the Ne\,{\sc x} K$\alpha$ line (1.0 keV).
With iron at five times solar abundance, the increased 
iron absorption lowers the continuum for $1\la E\la 40\,{\rm keV}$, 
while the iron K$\alpha$ (6.7 keV), L$\alpha$ ($\sim\!1\,{\rm keV}$), and 
$2{\rm p}\rightarrow 2{\rm s}$ ($\sim\!0.1\,{\rm keV}$) emission lines are 
enhanced. At the same time, the K$\alpha$ lines of Mg, Si, and S are 
weakened, partly because these elements do not remain highly ionized to
as great a depth.

The grid of 462 models is being made available as an ionized-reflection
model, called {\sc reflion}, for use in the XSPEC data-fitting routine
(Arnaud 1996).

\section{Discussion}
The treatment of Ross \& Fabian (1993) has been extended to include
all commonly important ionization states and transitions. The
resulting spectral grids should be useful for interpreting spectra
from AGN, BHC and GRB. Major signatures of moderately ionized
reflection are a Compton hump at about 30~keV, a strong ionized iron
line at 6.7~keV, multiple emission lines on top of a soft continuum
between 0.3 and 3~keV and a further EUV hump of emission peaking at
about 60~eV. The iron abundance has a significant effect on the shape
of the reflected continuum between about 3 and 30~keV.

\begin{figure}
\includegraphics[scale=0.35,angle=270]{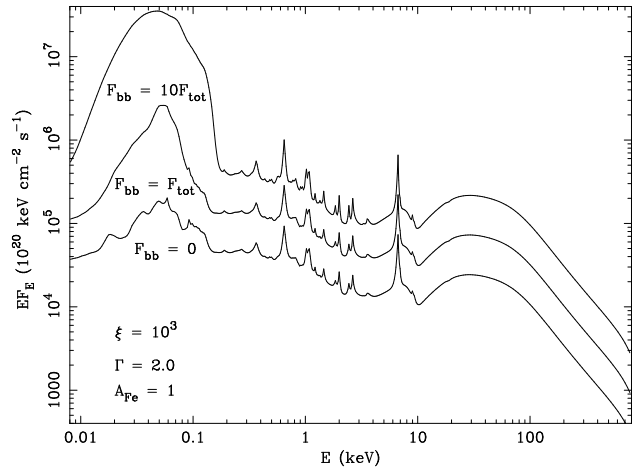}
\caption{Emergent spectra for pure reflection (bottom curve) and when
a soft blackbody spectrum ($kT_{\rm bb}=0.010$ keV) enters the surface 
layer from below. Results are shown with the blackbody flux equal to 
the illuminating flux (middle curve) and ten times greater than the
illuminating flux (top curve). The illuminating spectrum has 
$\xi=10^3$ erg\,cm\,s$^{-1}$ and $\Gamma = 2.0$, and iron has solar 
abundance. Successive spectra have been offset by factors of three 
for clearer presentation.}
\end{figure}

The spectra that we have calculated result entirely from 
reflection (reprocessed illumination). A significant fraction of the
incident X-ray energy emerges as softer radiation (as discussed recently
by Madej and R\'{o}\.{z}a\'{n}ska 2004, for example). The accretion 
disc beneath the illuminated surface layer can be expected to produce 
additional soft radiation (the ``big blue bump'') that can enhance the EUV 
hump. Figure~6 shows what happens when the calculations include a soft 
blackbody spectrum with $kT_{\rm bb}=0.010$ keV entering the illuminated 
layer from below. Models are shown with blackbody fluxes equal to the 
illuminating flux and ten times greater than the illuminating flux.
In both cases, more EUV radiation emerges than for pure reflection, as
expected, but the X-ray spectrum is essentially unchanged. Furthermore, 
accretion power liberated in an accretion disc corona (producing
the illumination) should reduce the amount of power liberated locally
within the disc, so the ``big blue bump'' may not be produced in the
same regions of the disc as the X-ray reflection. For these reasons, we 
have chosen to calculate reflection spectra only.

A signature of ionized reflection is the soft excess emission which
occurs in the 0.2-2~keV band due to lines and bremsstrahlung
from the hot surface layers. It gives a bump in the
relativistically-blurred spectrum which, when folded though an
XMM-Newton pn response matrix in order to simulate real data, is
well-fitted by a blackbody of temperature 150~eV (residuals less than
10 per cent). This may be relevant to the 100-200~eV temperature
component found in the X-ray spectra of many low-redshift PG quasars
(Gierlinski \& Done 2004; Porquet et al.\ 2004) which therefore
may be due to ionized reflection.

\begin{figure}
\includegraphics[width=0.32\textwidth,angle=270]{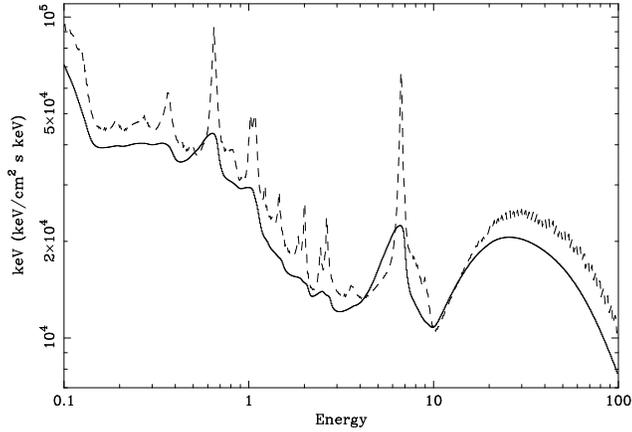}
\caption{Effect of relativistic blurring on the emergent spectrum. The
model with $\xi=10^3$ erg\,cm\,s$^{-1}$, $\Gamma=2.0$, and $A_{\rm Fe}=1$
(central line in Fig.~3, now dashed) has been
blurred as if observed from a disc with emissivity index 3 extending 
from 3 to 100 gravitational radii around a maximal Kerr black hole, 
viewed at an inclination angle of 30 deg (solid line).}
\end{figure}

We have checked whether the many emission features in the reflection
spectrum are likely to be readily observable. If there is little to
blur the spectrum as a result of either instrumental resolution or
shear within the ionized reflector, then the lines will be observable.
However, when there is strong shear, as in the inner parts of an accretion
disc, then the situtaion is much more difficult. We show in Figure~7 the
effect of relativistic blurring in an accretion disc inclined at 30
deg where the emission has an emissivity index of 3 and extends from 3
to 100 gravitational radii. Here $\xi=10^3$ erg\,cm\,s$^{-1}$, 
$\Gamma = 2.0$, and $A_{\rm Fe}=1$. The broad
(helium-like) iron-K line is very clear, as is a relatively weak broad
oxygen line. A weak feature of Fe-L and Ne with about the same
equivalent width as O ($\sim 120$~eV relative to the local reflection
`continuum' alone), together with Si and S are just discernible in
this reflection-only spectrum. Figure~8 demonstrates how extreme
relativistic blurring and changes in iron abundance affect
observations of iron-K features. A low iron abundance, perhaps due to
SN\,II enrichment, produces a small edge in the spectrum, whereas a
high iron abundance, perhaps due to SN\,Ia enrichment, produces an
apparent large edge.

\begin{figure}
\includegraphics[width=0.32\textwidth,angle=270]{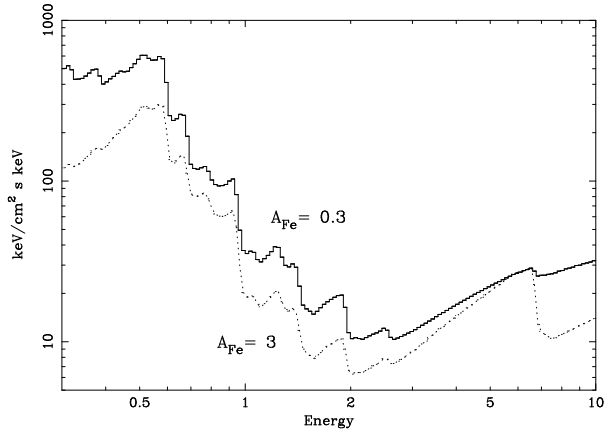}
\caption{Effects of iron abundance on relativistic-blurred
spectra.  Models for $\xi=30$ erg\,cm\,s$^{-1}$ and $\Gamma=2.0$
with $A_{\rm Fe}=0.3$ (solid) and $A_{\rm Fe}=3$ (dashed) have been 
blurred as if observed from a disc with emissivity index 3 extending 
from 2 to 100 gravitational radii around a maximal Kerr black hole, 
viewed at an inclination angle of 30 deg.}
\end{figure}

\section{Acknowledgements}
RRR and ACF thank the College of the Holy Cross and the Royal
Society, respectively, for support.


\bsp 

\label{lastpage}

\end{document}